\documentclass[aps,prl,twocolumn,groupedaddress]{revtex4}

\usepackage{graphicx} 
\usepackage{dcolumn}
\usepackage{bm}

\begin{document}
\title{Absence of nodes in the energy gap of the high-temperature
 ÊÊÊ electron-doped Pr$_{1.85}$Ce$_{0.15}$CuO$_{4-y}$ superconductor using thermal
 ÊÊÊ conductivity and specific heat measurements } 
\author{Guo-meng Zhao$^{1,2,*}$} 
\affiliation{$^{1}$Department of Physics and Astronomy, 
California State University, Los Angeles, CA 90032, USA~\\
$^{2}$Department of Physics, Faculty of Science, Ningbo
University, Ningbo, P. R. China}
\begin{abstract}
    We present data analyses of the thermal conductivity and specific
heat of electron-doped
Pr$_{1.85}$Ce$_{0.15}$CuO$_{4-y}$. The zero-field thermal conductivity and specific
heat of this
optimally electron-doped system
can be only explained by a nodeless gap symmetry. The
magnetic-field
dependence of the electronic specific heat in the vortex state is in quantitative
agreement with an $s$-wave theory.
Our quantitative data analyses provide bulk evidence for a nodeless gap symmetry 
in optimally electron-doped cuprates.

\end{abstract}

\maketitle

\vspace{0.3cm}



An unambiguous determination of the gap symmetry in the bulk of cuprates is 
crucial to the understanding of 
the pairing mechanism of high-temperature superconductivity. Three major gap
symmetry contenders have been $s$-wave,
$d$-wave, and extended $s$-wave ($s+g$ wave).  Both $d$-wave and extended $s$-wave have line nodes 
and change signs when a node is crossed.  The majority of measurements
probing low-energy excitations in the superconducting state (e.g.,
magnetic penetration depth, thermal conductivity, and 
specific heat) have pointed towards the existence of line 
nodes in the gap function of hole-doped cuprates \cite{Hardy,Jacobs,Lee,Chiao}.  
Qualitatively, these experiments are consistent with both $d$-wave and 
extended $s$-wave gap symmetries. In contrast to hole-doped 
cuprates, the earlier Raman scattering
\cite{Stad} and magnetic 
penetration depth data \cite{Alff} suggested a nodeless $s$-wave gap
symmetry in optimally electron-doped ($n$-type) cuprates. Later on, phase and surface-sensitive
experiments \cite{Tsu,Ari} provide evidence for pure
$d$-wave symmetry in optimally doped and overdoped $n$-type cuprates.
Surface-sensitive angle-resolved
photoemission spectroscopy (ARPES) \cite{Arm01,Matsui} 
also supports $d$-wave gap symmetry for optimally electron-doped cuprates. 
On the other hand, nearly bulk-sensitive point-contact tunneling spectra
\cite{note}
show no zero-bias conductance 
peak (ZBCP) in optimally doped and overdoped samples
\cite{Bis,Qaz,Shan05,Shan08} while the ZBCP is
seen in the tunneling spectra of a deeply underdoped $n$-type cuprate 
with $T_{c}$ = 13 K (Ref.~\cite{Bis,Qaz}). 
These tunneling experiments indicate a possible crossover from $d$ wave in
deeply underdoped materials to nodeless $s$ wave in optimally 
doped and overdoped materials. Very recently, the bulk-sensitive Raman
scattering data of Nd$_{1.85}$Ce$_{0.15}$CuO$_{4-y}$
can be quantitatively explained by an anisotropic $s$-wave gap with
a large minimum gap of about 3 meV (Ref.~\cite{Zhaopreprint}), in
quantitative agreement with the earlier
penetration depth data \cite{Alff,Zhao01}. Further, the bulk- and
phase-sensitive data of impurity pair-breaking effect in optimally-doped
Pr$_{1.855}$Ce$_{0.145}$CuO$_{4-y}$ unambiguously rule out any $d$-wave
gap symmetry \cite{Zhaopreprint}. Therefore, the surface-sensitive experiments 
probe a $d$-wave gap at any doping level 
while the bulk-sensitive experiments see an $s$-wave gap in
optimally doped and overdoped $n$-type cuprates. This apparent contradiction
can be naturally resolved if the surfaces are deeply
underdoped so that the gap symmetry is $d$-wave. Experiments on
hole-doped cuprates \cite{Bet,Mann} indeed show that surfaces
and interfaces are significantly underdoped.

Here we provide quantitative analyses of the bulk-sensitive thermal conductivity and specific
heat data of optimally doped $n$-type Pr$_{1.85}$Ce$_{0.15}$CuO$_{4-y}$. We find that the zero-field thermal conductivity and specific
heat of this
optimally electron-doped system
can be only explained by a nodeless gap symmetry. The
magnetic-field
dependence of the electronic specific heat in the vortex state is in quantitative
agreement with an $s$-wave theory.
The thermal properties of the optimally electron-doped cuprates consistently provide 
bulk evidence for a nodeless 
$s$-wave gap symmetry.

 The low-temperature electronic thermal
 conductivity $\kappa_{el}$ and specific heat
$C_{el}$ can be used to distinguish between a nodeless gap function
 and a gap function with line nodes. Specific heat experiments are insensitive to the phase of the 
 gap, but can provide bulk information on the behavior
of the density of states $N(E)$ near the Fermi
level $E_{F}$. The quantity $C_{el}/T$  is proportional to
$N(E)$ averaged over an interval $k_{B}T$ around $E_{F}$. For
a $d$-wave or extended $s$-wave symmetry, the gap vanishes on lines of nodes
on the Fermi surface, therefore $N(E)$ averaged over all
directions in the reciprocal space behaves as $N(E)$ $\propto$ $|E|$. This implies $C_{el}/T$
$\propto$ $T$ for $T$ $<<$ $T_{c}$ in zero magnetic field. In contrast, for a
fully gapped superconductor, $C_{el}/T$ $\simeq$ 0 under the same
conditions. Similarly, the quantity $\kappa_{el}$ $\propto$ $T$ at $T$ $<<$ $T_{c}$ in zero magnetic field
for $d$-wave or extended $s$-wave gap symmetry while for a
fully gapped superconductor, $k_{el}/T$ $\simeq$ 0 under the same
conditions.

More specifically, the 
 quantity $\kappa_{el}/T$ in zero magnetic-field is directly related to the Fermi velocity 
 $v_{F}$ and momentum $\hbar k_{F}$ in the nodal directions, and to the slope $S 
 = d\Delta (\theta$)/d$\theta$ at nodes, where $\Delta (\theta$) is 
 a gap function and $\theta$ is the angle measured from the Cu-O
 bonding direction.  The former two quantities can 
 be obtained from angle-resolved photoemission spectroscopic data while the latter one can be readily calculated 
 from the gap function.  The residual thermal 
 conduction is due to a fluid of zero-energy quasiparticles induced by 
 the pair-breaking effect of impurity scattering near the nodes in the 
 gap.  Calculations for the heat transport by nodal quasiparticles in 
 two dimensions give a general expression \cite{Chiao}
\begin{equation}\label{TM}
\frac{\kappa_{el}}{T}=\frac{k_{B}^{2}}{3\hbar}\frac{n}{d}(\frac{v_{F}}{v_{2}}+\frac{v_{2}}{v_{F}}),
\end{equation}
where $n/d$ is the stacking density of CuO$_{2}$ planes, $v_{2} = 
S/(\hbar k_{F})$.  One
can readily show that $S$ = 2$\Delta_{M}$ for a
simple $d$-wave gap function:  $\Delta (\theta) = \Delta_{M}\cos
2\theta$ (where $\Delta_{M}$ is the gap along the Cu-O bonding
directions). Similarly, the
low-temperature electronic specific heat in zero magnetic-field is proportional to $T^{2}$,
{\em i.e.}, $C_{el}$ = $\alpha T^{2}$, where $\alpha$ is a constant.

Another way to test the gap symmetry is to study the electronic
specific heat in the vortex state. According to
a simplified argument \cite{Vol}, the energy of carriers circulating
around a vortex is shifted by the Doppler effect. This shift
has a dramatic effect on the density of states when the gap is small, so that the essential 
contribution comes from the vicinity
of the nodes. In the low temperature limit, the density of states at
$E_{F}$ becomes proportional to $\sqrt{H}$ for one vortex so that
$C_{el}/T$ $\propto$ $\sqrt{H}$.  In an
isotropic $s$-wave superconductor, no significant
contribution should be expected at low temperatures from
such a mechanism, but localized states in vortex cores contribute a
term which is proportional to $H/H_{c2}$, that is, $C_{el}/T$ $\propto$ $H$. However, the above simple
picture should be valid only near the lower critical field $H_{c1}$
(Ref.~\cite{Ich}).
By taking account of  the
core excitations from the bound states in a vortex
core and the quasiparticle transfer between vortices, Ichioka {\em et 
al.} \cite{Ich} showed  $C_{el}(T,H)/T$ for an isotropic $s$-wave
superconductor is proportional to $H^{0.67}$ for $H \leq H_{c2}$ and $T$ =
0 while $C_{el}(T,H)/T$ for a $d$-wave superconductor is approximately 
proportional to $\sqrt{H}$ only for $H$ $<$ 0.1$H_{c2}$ and $T$ =
0.

\begin{figure}[htb]
    \includegraphics[height=13cm]{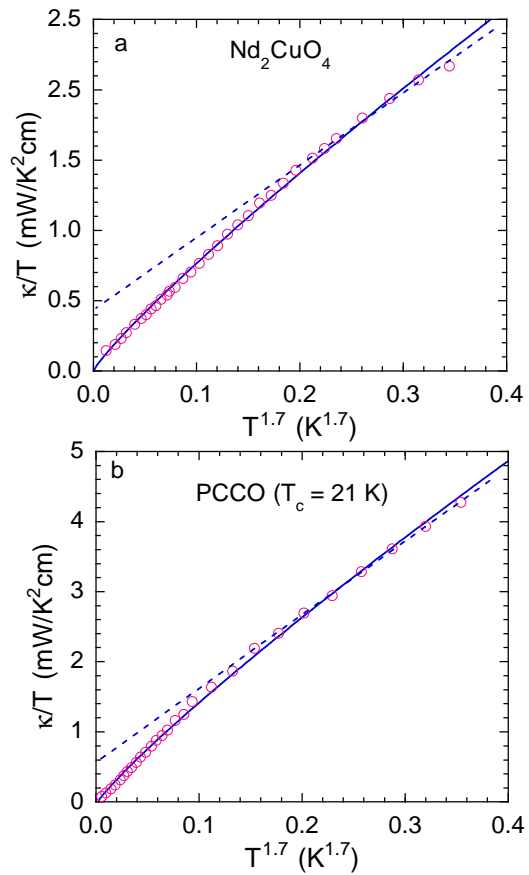}
 \caption[~]{a) $\kappa/T$ as a function of $T^{1.7}$ for an insulating
 parent compound  Nd$_{2}$CuO$_{4}$.  The data are 
from Ref.~\cite{Li}. The solid line is the fitted curve by an
emperical formula: $\kappa /T = A + BT^{p}$ with $A$ = 0 and $p$ =
1.49. The fit to the data above 0.38 K (dashed line) gives
$A$ = 0.45 mW/K$^{2}$cm and
$p$ = 1.7.  b) $\kappa/T$ as a function of $T^{1.7}$ for
for electron-doped
Pr$_{1.85}$Ce$_{0.15}$CuO$_{4-y}$ ($T_{c}$ = 21 K) at zero magnetic
field. The data are 
from Ref.~\cite{Rick}. The solid line is the fitted curve by 
the formula: $\kappa /T = A + BT^{p}$ with $A$ = 0 and $p$ =
1.49. The fit to the data above 0.33 K (dashed line) gives
$A$ = 0.50 mW/K$^{2}$cm and
$p$ = 1.7. }
\end{figure}

Figure~1a shows $\kappa/T$ as a function of $T^{1.7}$ for an insulating
parent compound  Nd$_{2}$CuO$_{4}$. The data are 
from Ref.~\cite{Li}. Low-temperature thermal conductivity data of several superconductors have 
been fitted by an
emperical formula: $\kappa /T = A + BT^{p}$, where $A$ and $B$ are
constants, and the exponent $p$ $\leq$ 2 (Ref.~\cite{Suth}). The constant $A$ is
equal to $\kappa_{el}/T$ in a $d$-wave superconductor and $A$ = 0 for 
a gapped $s$-wave superconductor. For the insulating material
Nd$_{2}$CuO$_{4}$ where $\kappa_{el}$ = 0, so the phonon contribution 
up to 0.5 K can be fitted by the emperical formula with $A$ = 0 and
$p$ = 1.49 (solid line). Fitting the limited data above 0.38 K (dashed 
linear line) gives
$A$ = 0.45 mW/K$^{2}$cm and
$p$ = 1.7. The inferred large value of $A$ = 0.45 mW/K$^{2}$cm
for this insulating compound is unphysical, indicating that the
fit to the limited data points leads to a wrong conclusion.

Figure~1b shows $\kappa/T$ as a function of $T^{1.7}$ for electron-doped
Pr$_{1.85}$Ce$_{0.15}$CuO$_{4-y}$ ($T_{c}$ = 21 K) in zero magnetic
field. The data are 
from Ref.~\cite{Rick}. Since the phonon contribution $\kappa_{ph}/T$ of the related
compound Nd$_{2}$CuO$_{4}$ follows a single 
power law with the exponent of 1.49 in the whole temperature range
from 0.07 K to 0.5 K, one should fit the data by the equation: $\kappa /T = A +
BT^{1.49}$.  The best fit (solid line) leads to $A$ = $\kappa_{el}/T$
= $-$0.036$\pm$0.015 mW/K$^{2}$cm. The negligibly small value of $\kappa_{el}/T$
should provide compelling evidence for a nodeless 
gap symmetry, in agreement with other independent  bulk-sensitive
experiments mentioned above. On the other hand, the fit to the data above 0.33 K
(dashed linear line) 
yields $A$ = 0.50 mW/K$^{2}$cm and $p$ = 1.7, which appears to be
consistent with $d$-wave gap symmetry. Since the fit to the limited data points 
of the insulating Nd$_{2}$CuO$_{4}$ leads to an unphysical
conclusion, the similar fit to the limited data points 
of the superconducting Pr$_{1.85}$Ce$_{0.15}$CuO$_{4-y}$ may also give
rise to a wrong conclusion.

\begin{figure}[htb]
    \includegraphics[height=13cm]{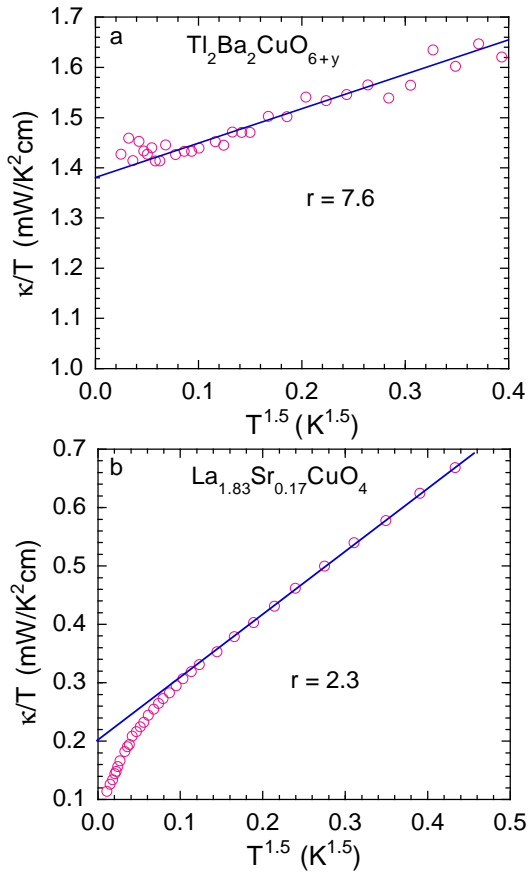}
 \caption[~]{a) $\kappa/T$ as a function of $T^{1.5}$ for a heavily overdoped Tl$_{2}$Ba$_{2}$CuO$_{6+y}$
($T_{c}$ = 15 K).  The data are 
from Ref.~\cite{Pro}. Since $\kappa_{ph}/T$ is proportional to
$T^{1.5}$ (see Fig.~1a above), we plot $\kappa/T$ as a function of $T^{1.5}$ to clearly see
if there is a low-$T$ downturn in electronic thermal conductivity.
Following the definition of the parameter $r$ in Ref.~\cite{Smith}, we
calculate $r$ = 7.6 using the reported sample dimensions (the length = 0.3 mm, the cross-sectional area 
= 0.2 mm$\times$0.01 mm) \cite{Pro}, and using the electronic contact
resistance = 0.2 $\Omega$ (Ref.~\cite{Pro}) and $\kappa_{el}/T$ = 1.39 mW/K$^{2}$cm.  b) $\kappa/T$ as a function of
$T^{1.5}$ for
for an overdoped La$_{1.83}$Sr$_{0.17}$CuO$_{4}$. The data are 
from Ref.~\cite{Haw}. We
calculate $r$ = 2.3 using the reported sample dimensions (the length =
0.97 mm, the cross-sectional area 
= 1.26 mm$\times$0.212 mm) \cite{Smith}, and using the electronic contact
resistance = 0.01 $\Omega$ (Ref.~\cite{Smith}) and $\kappa_{el}/T$ = 0.5 mW/K$^{2}$cm. }
\end{figure}

There are some other possible explanations for the negligibly small residual
linear term within the context of $d$-wave gap symmetry. The first
possibility is that $d$-wave nodal quasiparticles become completely
localized below 0.5 K.
However, the data in the magnetic field of 13 T indicate \cite{Hill} that the
complete localization would occur well below
0.1 K since $\kappa_{el}/T$ is still substantial (0.28~mW/K$^{2}$cm) at 0.1 K. This apparent contradiction makes this interpretation very
unlikely.  The second possibility is that the measurement
gradually ceases to detect the electronic heat current as electrons
fall out of thermal equilibrium with the phonon bath when the
temperature is below 0.5 K. Two facts lead us to rule out the electron-phonon 
decoupling \cite{Smith} as the main mechanism for the drop. First,
identical data are obtained by sending heat directly through the
electron system using photons (with no thermal resistance) \cite{Hill}. Secondly, 
if this model \cite{Smith} were relevant, the downturn
should be more pronouned in heavily overdoped Tl$_{2}$Ba$_{2}$CuO$_{6+y}$
($T_{c}$ = 15 K) than the overdoped La$_{1.83}$Sr$_{0.17}$CuO$_{4}$. This is
because the parameter $r$ that sensitively controls the extent of the 
low-$T$ downturn \cite{Smith} is calculated to be 7.6 for heavily overdoped
Tl$_{2}$Ba$_{2}$CuO$_{6+y}$, which is a factor of 
3.3 larger than that (2.3) for La$_{1.83}$Sr$_{0.17}$CuO$_{4}$. However, there is no
low-$T$ downturn in Tl$_{2}$Ba$_{2}$CuO$_{6+y}$ (Fig.~2a) while the significant drop starts
below about 0.26 K in La$_{1.83}$Sr$_{0.17}$CuO$_{4}$ (Fig.~2b).
Therefore, it is hard to imagine that this model would  be valid only for La$_{1.83}$Sr$_{0.17}$CuO$_{4}$
and Pr$_{1.85}$Ce$_{0.15}$CuO$_{4-y}$ but not for Tl$_{2}$Ba$_{2}$CuO$_{6+y}$.

\begin{figure}[htb]
    \includegraphics[height=13cm]{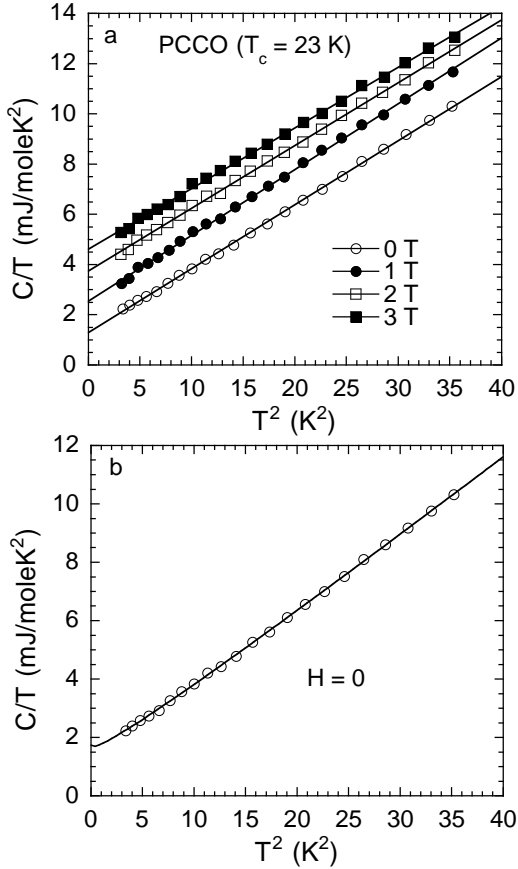}
 \caption[~]{a) Temperature dependence of specific heat 
for a Pr$_{1.85}$Ce$_{0.15}$CuO$_{4-y}$ crystal ($T_{c}$ = 23 K), at different
magnetic fields ($H$ $\parallel$ $c$). The data are taken from
Ref.~\cite{Ham}. The solid lines are linearly fitted curves.  b) Temperature dependence of 
specific heat at zero
magnetic field. The data are fitted by $C(T,0)/T =  \gamma
(0,0) + \alpha T + \beta T^{2}$.  The fitted parameter $\alpha$ is negative ($-$0.25$\pm$0.05 mJ/mol
K$^{3}$) and unphysical, suggesting that the
value of $\alpha$ is negligibly small, in agreement with a nodeless gap symmetry. }
\end{figure}

In Figure~3a, we show the temperature dependence of specific heat 
for a Pr$_{1.85}$Ce$_{0.15}$CuO$_{4-y}$ crystal ($T_{c}$ = 23 K), at different
magnetic fields ($H$ $\parallel$ $c$). The data are taken from
Ref.~\cite{Ham}. 
If the optimally doped cuprate is a gapped $s$-wave superconductor,
$C(T,H)/T$ should be expressed as
$C(T, H)/T =  \gamma (0, H) + \beta T^{2}$, where the first term is the
field-dependent electronic specific heat and the second term is the lattice specific heat 
due to acoustic phonons. It is clear that the specific heat data at
different magnetic fields can be well fitted by the above expression. 
The average of the fitted $\beta$ values is 0.25 mJ/mol K$^{4}$,
leading to a Debye temperature of 384 K. A finite value of $\gamma
(0, 0)$ is also found in hole-doped cuprates, and there are several
proposals to explain its origin \cite{Hus,Kre,Phi}.

On the other hand, if electron-doped cuprates are $d$-wave
superconductors with line nodes,
$C(T, 0)/T$ should be 
expressed as $C(T, 0)/T =  \gamma (0, 0) +
\alpha T + \beta T^{2}$, where the term $\alpha T$ arises from nodal
quasiparticle excitation (see discussion above). In Fig.~3b, we fit the zero-field data by 
$C(T, 0)/T =  \gamma
(0,0) +
\alpha T + \beta T^{2}$.  Although the three-parameter fit improves the 
quality of fitting slightly compared with the above two-parameter
fit, the fitted parameter $\alpha$ is negative ($-$0.25$\pm$0.05 mJ/mol
K$^{3}$) and unphysical. This suggests that the
value of $\alpha$ is negligibly small, in disagreement with pure $d$-wave gap symmetry
with line nodes.

\begin{figure}[htb]
    \includegraphics[height=6cm]{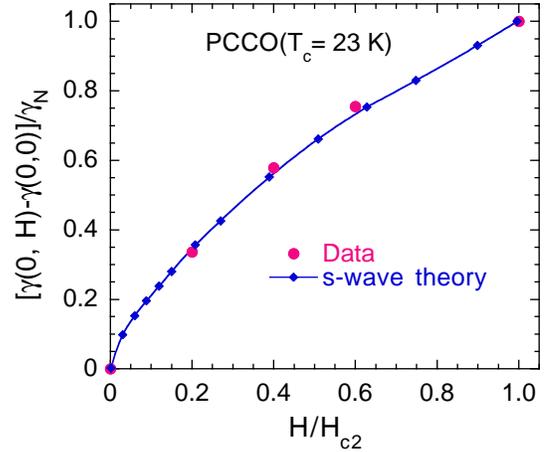}
 \caption[~]{Magnetic field dependence of $[\gamma (0, H) - \gamma 
(0, 0)]/\gamma_{N}$ for the Pr$_{1.85}$Ce$_{0.15}$CuO$_{4-y}$ crystal
(solid circles)
along with the theoretical prediction (solid diamonds) based on
isotropic $s$-wave gap symmetry \cite{Ich}. Here $H_{c2}$ and $\gamma_{N}$
for Pr$_{1.85}$Ce$_{0.15}$CuO$_{4-y}$
are taken to be 50 kOe and 4.1 mJ/mol K$^{2}$, respectively.}
\end{figure}

In order to obtain the zero-temperature $\gamma
(0, H)$ more precisely, we fit the data of Fig.~3a using $C(T, H)/T =  \gamma
(0,H) + \beta T^{2}$ with a fixed $\beta$ = 0.25 mJ/mol K$^{4}$.
Fig.~4 shows $[\gamma (0, H) - \gamma 
(0, 0)]/\gamma_{N}$ as a function of $H/H_{c2}$. Here $H_{c2}$ is
taken to be 50 kOe (see Fig.~2a of Ref.~\cite{Ham}) and $\gamma_{N}$ =
$[\gamma$(0,
56kOe)$- \gamma 
(0, 0)]$ = 4.1 mJ/mol K$^{2}$. It is striking that the field dependence of the electronic 
specific heat for this electron-doped cuprate is in quantitative
agreement with a theory based on isotropic $s$-wave gap symmetry
\cite{Ich}.
This excellent agreement clearly indicates that the optimally electron-doped 
cuprate is a gapped superconductor.

\begin{figure}[htb]
    \includegraphics[height=6cm]{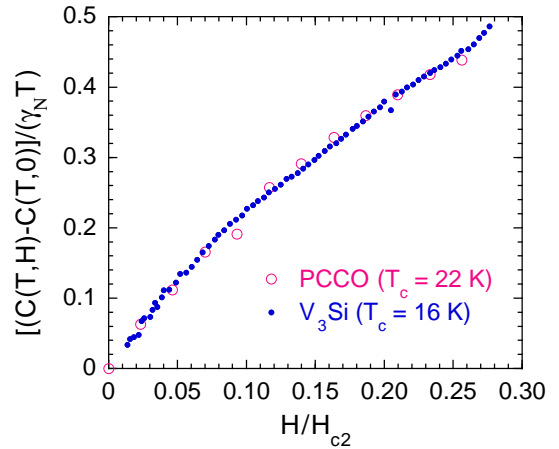}
 \caption[~]{Magnetic field dependence of $[C(T,H)-C(T,0)]/\gamma_{N}T$ 
 for a Pr$_{1.85}$Ce$_{0.15}$CuO$_{4-y}$ crystal
with $T_{c}$ = 22 K  at $T$ = 1.8 K (open circles) and for V$_{3}$Si at 
$T$ = 3.5 K (solid circles).  Fitting the two sets of data
respectively by a power law gives the
same power of 0.78. }
\end{figure}

Fitting the zero-temperature data and theoretically calculated values respectively in
Fig.~4 by a power law 
yields an exponent of 0.64$\pm$0.04 for the data and 0.663$\pm$0.008 for the theory. 
At finite temperatures, the exponent
should increase with increasing temperature, as demonstrated in an $s$-wave 
superconductor NbSe$_{2}$ (Ref.~\cite{Son}). In order to
further verify a nodeless gap symmetry in
the optimally electron-doped system, Fig.~5 compares $[C(T,H)-C(T,0)]/\gamma_{N}T$ for a Pr$_{1.85}$Ce$_{0.15}$CuO$_{4-y}$
crystal ($T_{c}$ = 22 K) with that for V$_{3}$Si.  The data of Pr$_{1.85}$Ce$_{0.15}$CuO$_{4-y}$
are reproduced from Ref.~\cite{Yu} and the data of V$_{3}$Si are from
Ref.~\cite{Ram}. The
values of $\gamma_{N}$ and $H_{c2}$ for the PCCO compound are taken to be 3.7
mJ/mol~K$^{2}$ 
and 43 kOe (see Fig.~2 of Ref.~\cite{Yu}), respectively. The
value of $H_{c2}$ for the V$_{3}$Si superconductor is found to be  250 kOe (Ref.~\cite{Ram}).
Using the
specific-heat jump at $T_{c}$ ({\em i.e.}, $\Delta C/T_{c}$ $\simeq$ 90 mJ/mol K$^{2}$
\cite{Ram}) and the
relation $\Delta C/\gamma_{N}T_{c}$ = 2.0 suitable for
V$_{3}$Si~\cite{Carb}, we calculate $\gamma_{N}$ to be 45 mJ/mol~K$^{2}$.  It is evident that the two sets of data almost
coincide with each other. Fitting the two sets of data respectively by a power law 
produces the
same exponent of 0.78, which is also very close to that (0.75) for NbSe$_{2}$ at
2.3 K (Ref.~\cite{Son}). Therefore, electron-doped cuprates have a very similar field 
dependence of the electronic specific heat as the conventional $s$-wave
superconductors.

In summary, independent bulk-sensitive specific heat and thermal
conductivity experiments show a nodeless gap symmetry in the bulk of optimally electron-doped cuprates. A $T^{2}$ dependence of the penetration depth  at low
temperatures observed in some samples \cite{Pr} can be well
explained \cite{Zhao01} in terms of nodeless $s$-wave gap along with an
extrinic effect due to current-induced nucleation of vortex-antivortex pairs at
defects. Although the data of spin-lattice relaxation rate and
Knight shift in an underdoped Pr$_{0.91}$LaCe$_{0.09}$CuO$_{4-y}$ are 
consistent with $d$-wave gap symmetry \cite{Zheng}), we will show that  
the data can be better explained by an anisotropic $s$-wave gap with a minimum
gap size being consistent with independent magnetic penetration depth data.
Further, the $d$-wave gap symmetry in Pr$_{0.91}$LaCe$_{0.09}$CuO$_{4-y}$ 
is incompatible with the observed large residual resistivity
(92~$\mu\Omega$cm) \cite{Zheng}, which would suppress $T_{c}$ to zero
\cite{Zhaopreprint}. Therefore, these bulk-sensitive experiments consistently point towards a nodeless 
$s$-wave gap symmetry, in agreement with predominantly phonon-mediated
pairing mechanism \cite{Huang,Zhao09}.

$^{*}$Correspondence should be addressed to gzhao2@calstatela.edu


\begin{thebibliography}{99}
\bibitem{note}The single-particle tunneling 
experiments can probe the bulk electronic density of states when
the mean free path is far larger than the thickness of the
degraded surface layer \cite{Pon}. Therefore, the single-particle tunneling
experiments along the CuO$_{2}$ planes are almost bulk sensitive due to a large in-plane mean free path
($>$ 100 \AA).
\end{thebibliography}
\end{document}